\documentstyle[12pt]{article}

         \def\la{\lambda}
         \def\be{\begin{equation}}
         \def\bea{\begin{eqnarray}}
         
         \def\o{\over}
         \def\ep{\epsilon}

         \def\ee{\end{equation}}
         \def\eea{\end{eqnarray}}
         \def\R{\rm {I\kern-.200em R}}
         \def\C{\rm {I\kern-.520em C}}
         \def\c{\chi}

\begin{titlepage}
\begin{document}
\vspace*{5mm}
\begin{center} {\Large \bf Green functions of 2-dimensional Yang-Mills\\
\vskip 0.35cm
theories on nonorientable surfaces}\\
\vskip 1cm
M.Alimohammadi$ ^{a,b}$ \footnote {e-mail:alimohmd@rose.ipm.ac.ir} and
M.Khorrami$ ^{a,b,c}$ \footnote {e-mail:mamwad@rose.ipm.ac.ir}\\
\vskip 1cm
{\it $^a$ Physics Department, University of Teheran, North Karegar,} \\
{\it Tehran, Iran }\\
{\it $^b$ Institute for Studies in Theoretical Physics and Mathematics, }\\
{\it P.O.Box 19395-5746, Tehran, Iran}\\
{\it $^c$ Institute for Advanced Studies in Basic Physics , P.O.Box 159 ,}\\
{\it  Gava Zang , Zanjan 45195 , Iran }\\
\end{center}
\vskip 2cm
\begin{abstract}

By using the path integral method , we calculate the Green functions of field
strength of Yang-Mills theories on arbitrary nonorientable surfaces in
Schwinger-Fock gauge. We show that the non-gauge invariant correlators consist
of a free part and an almost $x$-independent part. We also show that the gauge
invariant $n$-point functions are those corresponding to the free part , as in
the case of orientable surfaces.

\end{abstract}
\newpage

It has been long known that the two-dimensional Yang-Mills theory is exactly
soluble and indeed locally trivial [1]. The reason for this , at a fundamental
level , is that in the two dimensions the Yang-Mills action depends only on
the measure $\mu$ determined by the metric $g$. In the recent years these
theories has been studied more carefully again.

In refs.[2] and [3] the partition function and the expectation value of Wilson
loops has been calculated by means of lattice gauge theory for arbitrary
two-dimensional closed Riemann surfaces. These quantities also have been derived
in the context of path integral in [4] and [5] and the abelianization technique
were used to study this theory in refs.[6] and [7]. There are also some efforts
to calculate the explicit expression of the partition function on sphere in
small area limit [8,9] and the Wilson loops for $SU(N)$ gauge group [10].

The other interesting quantities that must be calculated are the Green functions
of field strength. In ref.[11] some of the correlators have been calculated
by the abelianization method and in ref.[12] all $n$-point functions have
been derived by path integral method for arbitrary closed orientable Riemann
surfaces. There we have shown that the gauge invariant Green functions correspond
to a free field theory.

In this paper we are going to complete our investigation about the correlators
of $2d$ Yang-Mills theories by calculating them on arbitrary closed nonorientable
surfaces. Such surfaces are connected sums of an orientable suface of genus $g$
with $s$ copies of Klein bottle and $r$ copies of the projective plane ;
$\Sigma_{g,s,r}$. $\Sigma_{g,s,r}$ can also be regarded as the connected sum
of $r+2(s+g)$ projective planes , provided $r$ and $s$ are not both zero.
The procedure that we follow are along that of ref.[12].

To begin , let us first rederive the partition function of Yang-Mills theories
on $\Sigma_{g,s,r}$ by path integral method. This quantity has been derived in
ref.[3] in the context of lattice gauge theory
by using the Migdal's suggestion about the local factor of plaquettes. First
we consider $r=0$ case. Consider a genus-$g$ Riemann surfaces with $n=2s$
boundaries. The boundary condition of each boundary loop $\gamma_i$ is
specified by a group element $g_i$ , such that ${\rm Pexp} \oint_{\gamma_i} A
=g_i\in G$ , where $A$ is the gauge field and $G$ is an arbitrary non-abelian
compact semisimple gauge group. We take the specific case of boundary condition
in which the $g_i$'s are $g_1,g_1^{-1},...,g_s,g_s^{-1}$. The wave function
corresponding to this situation is [4] :
\be \psi (\Sigma_{g,n=2s},g_1,..., g_s^{-1})=\sum_\la d(\la)^{2-2g-2s}
\c_\la (g_1) ... \c_\la (g_s^{-1}) e^{-{\ep \o 2}c_2(\la)A}.\ee
In this relation $\la $ labels the irreducible unitary representation of $G$ ,
$\c_\la $ is the character , $d(\la )$ the dimension and $c_2(\la )$ the
quadratic Casimir of the representation. $A$ is the area of $\Sigma_{g,n}$ and
$\ep $ is the coupling constant. Note that the Schwinger-Fock gauge was used
in calculation of the above wave function [4,12] :
\be A_\mu^a(x)=\int_0^1dssx^\nu F^a_{\nu \mu}(sx). \ee
In the remaining of this paper we will also work in this gauge.

Now if we integrate the eq.(1) over $g_1,...,g_s$ we will find the partition
function of genus $g+s$ orientable surface. But as was mentioned in [3] the
action of an orientation-reversing diffeomorphism on boundaries is :
\be \c_\la (g) \longrightarrow \c_\la (g^{-1})= \c_{\bar \la } (g) ,\ee
where ${\bar \la }$ is the complex conjugate representation of $\la $. So if we
change $\c_\la (g_i^{-1}) \longrightarrow \c_{\bar \la } (g_i^{-1})$ in eq.(1)
and then integrate over $g_i$'s , we will recover the partition function of
a genus-$g$ surface with $s$ copies of Klein bottle :
\be Z_{g,s,r=0}=\sum_{\la={\bar \la}} d(\la)^{2-2g-2s}e^{-{\ep \o 2}c_2(\la)A},\ee
which is the same as what obtained in [3] . In the above we have used the
following orthogonality relation of characters :
\be \int \c_\la (g) \c_\mu (g^{-1})dg=\delta_{\la \mu } , \ee

Now let us consider $RP^2$. Consider a disk with boundary
condition
\be {\rm Pexp} \oint_{\partial D} A=g=g'^2. \ee
The wave function of this disk is
\be \psi_D (g'^2)=\sum_\la d(\la) \c_\la (g'^2)e^{-{\ep \o 2}c_2(\la)A}.\ee
To obtain the partition function of $RP^2$ , we just have to integrate over $g'$
, so that we have glued the boundaries in such a manner to obtain $RP^2$ from
the disc. The partition function thus becomes :
\be Z_{RP^2}=\int dg'\psi_D(g'^2) =\sum_\la f_\la  e^{-{\ep \o 2}c_2(\la)A},\ee
where
\be f_\la:=\int \c_\la (g^2)dg. \ee
$f_\la=0$ unless the representation $\la$ is self conjugate.
If the representation $\la$ is self conjugate , there exist an invariant bilinear
form. Then $f_\la=1$ if this bilinear form is symmetric , and $f_\la =-1$ if it is
antisymmetric [13].

Now consider a sphere with $r'+n$ boundaries. If one turns $r'$ boundaries into
$RP^2$ , the connected sum of a sphere with $n$ boundaries and $r'=2g+2s+r$
$RP^2$'s is obtained. In this way the wave function is obtained to be :
$$ \psi (g_1,..., g_n)=\int dg'_1...dg'_{r'}\sum_\la d(\la)^{2-r'-n}
\c_\la (g'^2_1) ... \c_\la (g'^2_r) \c_\la (g_1) ... \c_\la (g_n)
e^{-{\ep \o 2}c_2(\la)A} $$
\be =\sum_\la f_\la^{r'}d(\la)^{2-r'-n} \c_\la (g_1)...\c_\la(g_n)
e^{-{\ep \o 2}c_2(\la)A}.\ee
Note that this surface is the most general nonorientable surface with boundaries
, because it is well known that a surface with $g$ handles , $s$ Klein bottles ,
and $r$ projective planes , is in fact nothing but a sphere with $r'=r+2(g+s)$
projective planes ( provided $r$ and $s$ are not both zero ). Comparing (10)
with (4) , shows this once again : (4) is a special case of (10) , with $n=0$
and $r'=2(g+s)$. To summarize , it is shown that
\be \psi (\Sigma_{g,s,r},g_1,..., g_n)=
\sum_\la f_\la^{r+2s}d(\la)^{2-2g-2s-r-n} \c_\la (g_1)...\c_\la(g_n)
e^{-{\ep \o 2}c_2(\la)A},\ee
and this relation holds for orientable , as well as nonorientable surfaces.

Now everything is ready to calculate the $n$-point functions of field strength
on $\Sigma_{g,s,r}$. It was shown in [12] that the wave function on a disk ,
with boundary condition $g_1$ , in the presence of a source function $J(x)$ and
in Scwhinger-Fock gauge is ( this equation can also be found in the appendix
of [11] )
$$ \psi_D[J]=\int {\cal D} \xi e^{-{1 \o 2\epsilon}\int \xi^a\xi_ad\mu+\int\xi^a J_a d\mu}
\delta({\rm Pexp}\oint_\gamma A,g_1)$$
\be = Z_1[J]\psi_{2,D}[J] ,\ee
where
\be Z_1[J]=e^{{\ep \o 2}\int J^aJ_ad\mu} ,\ee
and
\be \psi_{2,D}[J]=\sum_\la \c_\la (g_1^{-1})e^{-{\ep \o 2}c_2(\la)A(D)}
\c_\la ({\cal P}{\rm exp} \ep \int dt (\int ds {\sqrt g}J(s,t))) .\ee
$\xi (x)=\xi^a(x)T^a$ is defined by $F_{\mu \nu }=\xi (x) \ep_{\mu \nu},T^a$'s
are the generators of $G$ and $d\mu =\sqrt {g(x)}d^2x$ . The ordering in (14)
is according to $t$. the disk is parametrized by coordinates $s$ ( the radial
coordinate ) and $t$ ( the angle coordinate ). The functinal derivative of the
above wave functionals produce the Green functions , where for $\psi_D[J]$ is :
\be < \xi^{a_1}(x_1)...\xi^{a_n}(x_n)>_{2,D}
= \sum_\la \ep^n\c_\la (g_1^{-1})\c_\la (T^{a_1}...T^{a_n})
 e^{-{\ep \o 2}c_2(\la)A(D)}\ee
$ {\rm for} \ \ \ t(x_1)<...<t(x_n) $.

Now to calculate the $n$-point function of $\xi^a$'s on $\Sigma_{g,s,r}$ it is
enough to glue the expectation value (15) to the wave function (11) with $n=1$.
The result is :
$$ < \xi^{a_1}(x_1)...\xi^{a_n}(x_n)>_{2,\Sigma_{g,s,r}}={1 \o Z_{\Sigma_{g,s,r}}}
\int dg_1< \xi^{a_1}(x_1)...\xi^{a_n}(x_n)>_{2,D}\psi ( \Sigma_{g,s,r},g_1) $$
\be ={1 \o Z_{\Sigma_{g,s,r}}}
\sum_{\la} \ep^n f_\la^{r+2s}d(\la)^{1-2g-2s-r}\c_\la (T^{a_1} ... T^{a_n})
e^{-{\ep \o 2} c_2(\la)A}.\ee
The Green function corresponding to $Z_1$ , which is the partition function of
a free field theory , are [12] :
\be < \xi^{a_1}(x_1)...\xi^{a_{2n}}(x_{2n})>_{1,\Sigma_{g,s,r}}=\sum_p
G^{a_{i_{1}}a_{i_{2}}}(x_{i_1},x_{i_2})...
G^{a_{i_{2n-1}}a_{i_{2n}}}(x_{i_{2n-1}},x_{i_{2n}})\ee
where
\be G^{ab}(x,y)=\ep \delta^{ab}\delta (x-y) ,\ee
and the summation is over all distinct pairing of $2n$ indeces. The complete
$n$-point functions are :
\be<\xi (x_1) ... \xi (x_{n})>=\sum_{m=0}^{n}\sum_c<\xi (x_1) ... \xi (x_m)>_1
<\xi (x_{m+1}) ... \xi (x_{n})>_2,\ee
where the inner summation is over all different ways of choosing $m$ indeces
from $2n$ indeces. It is clear that the correlators consist of a free (eq.(17))
and an almost $x$-independent part (eq.(16)). This is the same behaviour that
was obsereved in the orientable case [12].

The last question that must be answered is that which part of the above Green
functions are gauge invariant. In ref.[12] , by two methods , we showed that
only the free part is gauge invariant. The first method was that if one constructs
the gauge invariant wave function on disk , $\psi_D^{G.I.}[J]$ , it can be shown
that their corresponding Green functions are those quoted in (17). The same
argument holds here and therefore in nonorientable case the gauge invariant
$n$-point functions correspond to a free field theory. The second method is to
calculate the Green functions by a different approach , that is by using the
expectation value of Wilson loops , which is gauge invariant. So we must first
calculate these expectation values. It is clear that it can be constructed as
follows
\be <\c_\mu ({\rm Pexp} \oint_\gamma A)> = {1 \o Z_{g,s,r}} \int \psi
(\Sigma_{g,s,r},g_1)\c_\mu (g_1) \psi_D (g_1^{-1})dg_1, \ee
where $\gamma = \partial D$ is a contractable loop. A simple calculation shows
\be <\c_\mu ({\rm Pexp} \oint_\gamma A)>={1 \o Z_{g,s,r}}\sum_\la \sum_{\rho \in
\la \otimes \mu}f_\la^{r+2s}d(\rho )d(\la )^{1-2g-2s-r} exp\{-{\ep \o 2} [c_2(\la )(A-A(D))+c_2(\rho )
A(D)] \} \ee
In small $A(D)$ limit , the LHS of (21) becomes
\be {\rm LHS }=d(\mu)+{1 \o 2}\int
\c_\mu <\xi (x) \xi (y) > d\mu (x) d\mu (y)+ ... \ee
Now if by symmetry consideration we use the following ansatz for the gauge invariant
two-point function
\be <\xi^a(x)\xi^b(y)>^{G.I.}=M\delta^{ab}\delta (x-y) , \ee
then eq.(22) reduces to
\be {\rm LHS }=d(\mu)-{M \o 2}A(D)d(\mu)c_2(\mu).\ee
In the same limit , it can be shown that the RHS of eq.(21) is [12]
\be {\rm RHS }=d(\mu)-{\ep \o 2}A(D)d(\mu)c_2(\mu),\ee
which shows that $M=\ep$ and is consistent with eq.(18) .

At the end , we must note that if the gauge group was $U(1)$ , we would produce
nothing for nonorientable surfaces , for the simple reason that non of the
representation of $U(1)$ are self conjugate except for the trivial representation.

{\bf Acknowledgement :} We would like to thank the research vice-chancellor
of the university of Tehran , which this work was partially supported by them.
\pagebreak

\end{titlepage}

\begin{thebibliography}{99}

\bibitem [1]{a} A. Migdal , Zh. Eskp. Teor. Fiz. 69 (1975) , 810 (Sov. Phys.
Jetp. 42 , 413 ).
\bibitem [2]{b} B. Rusakov , Mod. Phys. Lett. A5 (1990) , 693.
\bibitem [3]{c} E. Witten , Commun. Math. Phys. 141 (1991) , 153.
\bibitem [4]{d} M. Blau and G. Thompson , Int. Jour. Mod. Phys. A7 (1992) , 3781.
\bibitem [5]{e} G. Thompson , " Topological gauge theory and Yang-Mills theory "
, in Proceeding of the 1992 Trieste Summer School on High Energy Physics and
Cosmology , World Scientific , Singapore (1993) , 1-76.
\bibitem [6]{f} M. Blau and G. Thompson , Jour. Math. Phys. 36 (1995) , 2192.
\bibitem [7]{g} M. Blau and G. Thompson , Commun. Math. Phys. 171 (1995) , 639.
\bibitem [8]{h} D. Gross and A. Matytsin , Nucl. Phys. B429 (1994) , 50.
\bibitem [9]{i} M. Crescimanno , S. G. Naculich and H. J. Schnitzer , "
Evaluation of the free energy of two-dimensional Yang-Mills theory " ,
hep-th/9601104 .
\bibitem [10]{j} A. Ashtekar , J. Lewandowski , D. Marolf , J. Mourao and
T. Thiemann , " Closed formula for Wilson loops for $SU(N)$ quantum Yang-Mills theory
in two dimensions " , hep-th/9605128.
\bibitem [11]{m} J. P. Nunes and H. J. Schnitzer , " Field Strength Correlators
for two dimensional Yang-Mills Theories over Riemann Surfaces " , hep-th/9510154.
\bibitem [12]{k} M. Alimohammadi and M. Khorrami , " $n$-point functions of $2d$
Yang-Mills theories on Riemann surfaces " , Int. Jour. Mod. Phys. A (1996) (to
appear) , hep-th/9606071.
\bibitem [13]{l} T. Brocker and T. tom Dieck , " Representation of compact Lie
groups "  Berlin, Heidelberg, New York : Springer 1985.



\end{thebibliography}
\end{document}